# Inductive Proof of Borchardt's Theorem

Andy A. Chavez,[1] Alec P. Adam,[1] Paul W. Ayers,[2*] Ramón Alain Miranda-Quintana[1*]

1. Department of Chemistry, Quantum Theory Project, University of Florida, Gainesville, FL 32611, USA
2. Department of Chemsitry, McMaster University, Hamilton, ON L8S 4M1, Canada
Email: quintana@chem.ufl.edu, ayers@mcmaster.ca

*Abstract*

We provide an inductive proof of Borchardt's theorem for calculating the permanent of a Cauchy matrix via the determinants of auxiliary matrices. This result has implications for antisymmetric products of interacting geminals (APIG), and suggests that the restriction of the APIG coefficients to Cauchy form (typically called APr2G) is special in its tractability.

The antisymmetric nature of electrons (given by their fermionic character) is often perceived as an extra complication by those starting to learn about quantum chemistry.[1] However, we can seamlessly keep track of the required sign changes by using structures that guarantee the antisymmetry of many-fermion states like second quantization techniques or, equivalently, Slater determinants.[2,3] This is particularly convenient because determinants can be computed with great ease, which in part explains why methods based on the single-particle picture are so popular.[4-6] The problem with this approach is that when the electrons are strongly correlated it is not possible to single out a dominant Slater determinant.[7-9]

Among the many alternatives proposed to deal with strong correlation, one that has regained steam recently is the use of pairs of electrons (geminals) as the building blocks of the wavefunction.[10-24] Geminals effectively behave as hard-core bosons, since they cannot occupy the same state (like fermions) but, more importantly for us in this note, they commute (like bosons). This means that in this case it is more natural to deal with a different function of the elements of a matrix: the *permanent*. Permanents have all the same terms as determinants, but all of them only have positive signs. This might appear like a simplification, but from a computational point of view the calculation of a permanent is much more complicated than that of a determinant. While the latter scales polynomially with the size of the matrix (like $O(N^3)$), calculating a permanent is a #P-complex problem.[25-32] This implies that very accurate methods like the Antisymmetric

P.Product of Geminals (AGP) and the Antisymmetric Product of Interacting Geminals (APIG) are computationally unaffordable.[33-35] Popular strategies leading to tractable geminal wavefunctions entail imposing very strict restrictions on the structure of the geminals, like introducing occupied-virtual separations[36,37] (in the one reference orbital methods like the Antisymmetric Product of One-Reference Orbital Geminals, AP1roG)[38-41] or restricting geminals to subspaces of orbitals (in the antisymmetric product of strongly orthogonal geminals).[42-56] Recently, geminals inspired from exactly solvable models have become an attractive alternative.[17,57,58] One of their key advantages is that, due to the constraints imposed by the underlying physical models, the permanents of the geminal coefficient matrices can be computed efficiently. A key result in this regard is Borchardt's theorem, that expresses the permanent of a Cauchy matrix in terms of determinants.[17,59-61] This is a well-known result in combinatorics, that has been proven in different ways.[62] Here, we present an inductive proof of this theorem, which not only deepens our understanding of Borchardt's original result, but also provides a potential path to extending the result to other types of geminals.

What we want to prove is:

$$|\mathbf{C}||\mathbf{C}|^{+} - |\mathbf{C} \circ \mathbf{C}| = 0 \qquad (1)$$

where $|\mathbf{C}|, |\mathbf{C}|^{+}, |\mathbf{C} \circ \mathbf{C}|$ represent the determinant, permanent, and Hadamard (element-wise) square of matrix $\mathbf{C}$, which in turn is a Cauchy matrix, that is, with entries:

$$[\mathbf{C}]_{ij} = \frac{1}{x_i - y_j} \qquad (2)$$

Eq. (1) always holds for 1×1 and 2×2 matrices, so the first case we need to tackle is 3×3. (This is the correct starting point for a proof by induction, because every 2×2 matrix is, after row and column scaling, representable as a Cauchy matrix.) Let us briefly sketch the work for the 3×3 case, which elucidates the strategy for the inductive step:

$$\mathbf{C} = \begin{bmatrix} c_{11} & c_{12} & c_{13} \\ c_{21} & c_{22} & c_{23} \\ c_{31} & c_{32} & c_{33} \end{bmatrix} \qquad (3)$$

$$|\mathbf{C}|^{+} = c_{11}c_{22}c_{33} + c_{12}c_{23}c_{31} + c_{13}c_{21}c_{32} + \\ c_{13}c_{22}c_{31} + c_{11}c_{23}c_{32} + c_{12}c_{21}c_{33} \qquad (4)$$

$$|\mathbf{C}| = c_{11}c_{22}c_{33} + c_{12}c_{23}c_{31} + c_{13}c_{21}c_{32} + \\ -c_{13}c_{22}c_{31} - c_{11}c_{23}c_{32} - c_{12}c_{21}c_{33} \qquad (5)$$

$$|\mathbf{C} \circ \mathbf{C}| = (c_{11}c_{22}c_{33})^2 + (c_{12}c_{23}c_{31})^2 + (c_{13}c_{21}c_{32})^2 + \\ -(c_{13}c_{22}c_{31})^2 - (c_{11}c_{23}c_{32})^2 - (c_{12}c_{21}c_{33})^2 \tag{6}$$

Therefore:

$$|\mathbf{C}||\mathbf{C}|^+ - |\mathbf{C} \circ \mathbf{C}| = c_{11}c_{22}c_{33}c_{12}c_{23}c_{31} + c_{11}c_{22}c_{33}c_{13}c_{21}c_{32} + c_{13}c_{21}c_{32}c_{12}c_{23}c_{31} + \\ -c_{13}c_{22}c_{31}c_{11}c_{23}c_{32} - c_{13}c_{22}c_{31}c_{12}c_{21}c_{33} - c_{11}c_{23}c_{32}c_{12}c_{21}c_{33} \tag{7}$$

It is convenient to introduce the following notation:

$$d_{ij} \equiv \frac{1}{c_{ij}}; \quad D = \prod_{ij} d_{ij} \tag{8}$$

Now Eq. (7) can be rewritten as:

$$|\mathbf{C}||\mathbf{C}|^+ - |\mathbf{C} \circ \mathbf{C}| = D^{-1}\left[d_{11}(d_{22}d_{33} - d_{23}d_{32}) - d_{12}(d_{21}d_{33} - d_{23}d_{31}) + d_{13}(d_{21}d_{32} - d_{22}d_{31})\right] \\ = D^{-1}|\mathbf{D}| \tag{9}$$

This means that Borchardt's theorem will be valid if the determinant of the Hadamard inverse is equal to 0. Since we are working with 3×3 matrices the Hadamard inverse must be, at most, of rank 2 (which is trivially true for Cauchy matrices).

Now, to proceed with the proof by induction, let us assume that Borchardt's theorem is valid for the 3×3, 4×4,…,(N-1)×(N-1) cases, and let us consider the N×N case. We will have:

$$|\mathbf{C}_N| = \sum_{p=1}^{N!}(-1)^{1+p} c_{11_p}c_{22_p}...c_{NN_p} \tag{10}$$

$$|\mathbf{C}_N|^+ = \sum_{p=1}^{N!} c_{11_p}c_{22_p}...c_{NN_p} \tag{11}$$

here, $p$ is just a way to numerate the different permutations.

It is useful to rewrite the last equations as:

$$|\mathbf{C}_N| = \sum_{\sigma=2k+1}^{N!/2} c_{11_\sigma}c_{22_\sigma}...c_{NN_\sigma} - \sum_{\bar{\sigma}=2k}^{N!/2} c_{11_{\bar{\sigma}}}c_{22_{\bar{\sigma}}}...c_{NN_{\bar{\sigma}}} \tag{12}$$

$$|\mathbf{C}_N|^+ = \sum_{\sigma=2k+1}^{N!/2} c_{11_\sigma}c_{22_\sigma}...c_{NN_\sigma} + \sum_{\bar{\sigma}=2k}^{N!/2} c_{11_{\bar{\sigma}}}c_{22_{\bar{\sigma}}}...c_{NN_{\bar{\sigma}}} \tag{13}$$

Since:

$$\left(\sum_i a_i - \sum_i b_i\right)\left(\sum_i a_i + \sum_i b_i\right) = \sum_{ij} a_i a_j - \sum_{ij} b_i b_j \tag{14}$$

we only have to focus on the products of terms within each of the "sides" of Eq. (12): the "positive terms" (labeled by the $\sigma$ s, corresponding to even permutations) and the "negative terms" (labeled by the $\bar{\sigma}$ s, corresponding to odd permutations). Since:

$$|\mathbf{C}_N \circ \mathbf{C}_N| = \sum_{\sigma=2k+1}^{N!/2} \left(c_{11_\sigma} c_{22_\sigma} ... c_{NN_\sigma}\right)^2 - \sum_{\bar{\sigma}=2k}^{N!/2} \left(c_{11_{\bar{\sigma}}} c_{22_{\bar{\sigma}}} ... c_{NN_{\bar{\sigma}}}\right)^2 \quad (15)$$

we just have to show that:

$$\sum_{\sigma,\sigma'=2k+1} c_{11_\sigma}...c_{NN_\sigma} c_{11_{\sigma'}}...c_{NN_{\sigma'}} - \sum_{\bar{\sigma},\bar{\sigma}'=2k} c_{11_{\bar{\sigma}}}...c_{NN_{\bar{\sigma}}} c_{11_{\bar{\sigma}'}}...c_{NN_{\bar{\sigma}'}} = 0 \quad (16)$$

where $\sigma \neq \sigma', \bar{\sigma} \neq \bar{\sigma}'$.

The strings $c_{11_\sigma}...c_{NN_\sigma}$ and $c_{11_{\sigma'}}...c_{NN_{\sigma'}}$ could have, at most, *N*-3 terms in common: If they have *N*-1 terms in common then they will be equal. If they have *N*-2 terms in common, then they differ by a single permutation, and they would belong to permutations with different parity. Then, the products with common terms can be grouped as:

$$\sum_{r=1}^{N-3} \sum_p \sum_{\sigma,\sigma',\bar{\sigma},\bar{\sigma}'} c_{11_p}^2...c_{rr_p}^2 \left(c_{r+1r+1_\sigma}...c_{NN_\sigma} c_{r+1r+1_{\sigma'}}...c_{NN_{\sigma'}} - c_{r+1r+1_{\bar{\sigma}}}...c_{NN_{\bar{\sigma}}} c_{r+1r+1_{\bar{\sigma}'}}...c_{NN_{\bar{\sigma}'}}\right) \quad (17)$$

The key insight here is that the terms in parenthesis can be rewritten as:

$$\sum_{\sigma,\sigma',\bar{\sigma},\bar{\sigma}'} \left(c_{r+1r+1_\sigma}...c_{NN_\sigma} c_{r+1r+1_{\sigma'}}...c_{NN_{\sigma'}} - c_{r+1r+1_{\bar{\sigma}}}...c_{NN_{\bar{\sigma}}} c_{r+1r+1_{\bar{\sigma}'}}...c_{NN_{\bar{\sigma}'}}\right) = \\ \left|\mathbf{C}_{N-r}^{11_p...NN_p}\right| \left|\mathbf{C}_{N-r}^{11_p...NN_p}\right|^+ - \left|\mathbf{C}_{N-r}^{11_p...NN_p} \circ \mathbf{C}_{N-r}^{11_p...NN_p}\right| \quad (18)$$

Here, the matrix $\mathbf{C}_{N-r}^{11_p...NN_p}$ is the (*N-r*)×(*N-r*) sub-matrix that is obtained from matrix $\mathbf{C}_N$ when we eliminate the elements in rows and columns containing the elements $c_{11_p},...,c_{rr_p}$. But this is just Borchardt's theorem for these sub-matrices, so we will have:

$$\sum_{\sigma,\sigma',\bar{\sigma},\bar{\sigma}'} \left(c_{r+1r+1_\sigma}...c_{NN_\sigma} c_{r+1r+1_{\sigma'}}...c_{NN_{\sigma'}} - c_{r+1r+1_{\bar{\sigma}}}...c_{NN_{\bar{\sigma}}} c_{r+1r+1_{\bar{\sigma}'}}...c_{NN_{\bar{\sigma}'}}\right) = 0 \quad (19)$$

So, in Eq. (16) we need to keep track only of terms such that:

$$c_{11_p}...c_{NN_p} c_{11_{p'}}...c_{NN_{p'}}, \quad \forall k, k': kk_p \neq k'k'_{p'} \quad (20)$$

Finally, Borchardt's theorem follows from noticing that, for every term fulfilling Eq. (20) in the "even permutation" side there is going to be an identical term in the "odd permutation" side (which can be obtained rearranging the terms within the multiplication strings). Therefore, these terms cancel.

In reference 10, the Borchardt form was used to define the antisymmetric product of rank-2 geminals (APr2G) wavefunction form. An open question there was to what extent other forms (beyond a Cauchy matrix) would also lead to tractable electronic structure methods. A trivial extension is the multiplication of the Cauchy matrix by diagonal matrices to scale the rows and columns, using the identity $\left|\mathbf{DM\tilde{D}}\right|^{+} = \left|\mathbf{D}\right|^{+}\left|\mathbf{M}\right|^{+}\left|\mathbf{\tilde{D}}\right|^{+}$. There are known numerically tractable recursive methods for banded Hankel, Toeplitz, and Circulant matrices,[63-71] and the presence of those recursive techniques was partial motivation for this paper. These matrices have fewer free parameters than Cauchy matrices.

However, the general problem of finding structure matrices with numerically tractable permanents is extremely difficult, but also extremely important, as it has important applications in many areas of physics beyond geminal-product wavefunctions. Unfortunately, the recursive method here does not lend itself to nontrivial extensions beyond Cauchy matrices. The very first step of the recursion gives the *most general matrix whose Hadamard inverse is zero* (up to trivial row/column scaling), so the first step in the recursion would need to be an identity for the permanent of (perhaps a sub-class of) 3x3 matrices. Even if such a formula were hypothesized, an entirely different strategy would be needed for the inductive step, as the strategy here is strongly linked to the Borchardt form, Eq. (1). This was already hinted at in ref. 10, where it was noted that inverse-rank-*k* matrices (with $k \geq 2$) seemed to be numerically intractable. Based on this, and the known results for Hankel and Toeplitz matrices, we hypothesize that there are no numerically tractable algorithms for exactly (as opposed to probabilistically) numerically evaluating the permanents of *n*×*n* square matrices with *kn* parameters, where $k \geq 5$.

**Acknowledgements**

RAMQ thanks the Oak Ridge Associated Universities for a Ralph E. Powe Junior Faculty Enhancement Award. PWA acknowlges support from NSERC, the Canada Research Chairs, and the Digital Research Alliance of Canada.